\documentclass{eas}
\usepackage{graphicx}
\usepackage{epstopdf}
\usepackage{natbib}
\newcommand{\ha}{H\ensuremath{\alpha}}
\newcommand{\hi}{H~{\sc i}}
\newcommand{\vlsr}{\ensuremath{v_\mathrm{LSR}}}
\newcommand{\kms}{km s$^{-1}$}
\newcommand{\fdg}{\mbox{\ensuremath{.\!\!^\circ}}}
\newcommand{\arcdeg}{\mbox{\ensuremath{^\circ}}}%
\newcommand{\arcmin}{\mbox{\ensuremath{^\prime}}}%

\newcommand{\apj}{ApJ}
\newcommand{\apjl}{ApJ}
\newcommand{\apjs}{ApJS}

\newcommand{\aap}{A\&A}

\newcommand{\nat}{Nat}

\begin{document}
\bibliographystyle{astron}

\title{Ionization of Infalling Gas}
\thanks{WHAM and its research efforts are supported by National Science Foundation awards AST-0204973 and AST-0607512. AKD received support through REU site award AST-0453442. GJM acknowledges support from U. Sydney through a University Postdoctoral Fellowship.}
\author{L. M. Haffner}\address{Department of Astronomy, University of Wisconsin, Madison, WI, USA }
\author{A.K. Duncan}
\author{S.M. Hoffman}
\author{G.J. Madsen}\address{School of Physics, The University of Sydney, NSW 2006, Australia}
\author{A.S. Hill}\sameaddress{1}
\author{R.J. Reynolds}\sameaddress{1}
\begin{abstract}

\ha\ emission from neutral halo clouds probes the radiation and hydrodynamic conditions in the halo. Armed with such measurements, we can explore how radiation escapes from the Galactic plane and how infalling gas can survive a trip through the halo. The Wisconsin H-Alpha Mapper (WHAM) is one of the most sensitive instruments for detecting and mapping optical emission from the ISM. Here, we present recent results exploring the ionization of two infallling high-velocity complexes. First, we report on our progress mapping \ha\ emission covering the full extent of Complex A. Intensities are faint ($<100$ mR; EM $<0.2$ pc cm$^{-6}$) but correlate on the sky and in velocity with 21-cm emission. Second, we explore the ionized component of some Anti-Center Complex clouds studied by \citet{PeePutMcK07} that show dynamic shaping from interaction with the Galactic halo.

\end{abstract}
\maketitle

\section{Complex A}

\begin{figure}

\begin{center}
\includegraphics[width=9.5cm]{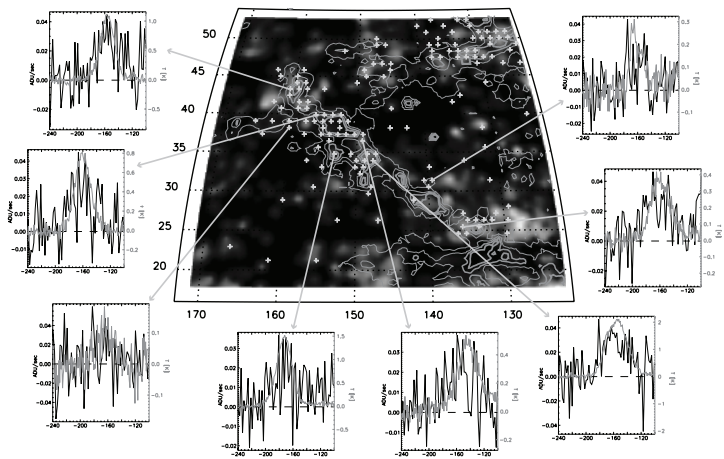}
\end{center}
\caption{
\label{compa}
Overlay of 21-cm contours (1, 3, 5, and $7 \times 10^{19}$ cm$^{-2}$; LDS, $0\fdg5$ beam) on an \ha\ intensity image (black, $< 20$ mR, white, $> 90$ mR; WHAM, $1\arcdeg$ beam) in the direction of Complex A. Axes are in Galactic degrees. White crosses show locations where means of fitted neutral and ionized emission components are close in velocity ($\Delta v < 10$ \kms). Spectra show example profiles from some of these directions (\ha, black; 21-cm, gray).
}

\end{figure}

HVC Complex A is a distant ($> 8$ kpc, \citealt{WakSavSem03}; $< 10$ kpc, \citealt{vanSchPel99}) collection of infalling neutral gas ($\vlsr \sim -200$ to $-130$ km s$^{-1}$). Its metallicity and origin are still uncertain, although a detection of O~{\sc i} by \citet{KunLeqSar94} suggests neutral oxygen is $\leq 0.1$ solar \citep{Wak01}. Figure~\ref{compa} shows the 21-cm emission from the complex.
\cite{TufReyHaf98} provides the first measurement of \ha\ in the complex toward cores A III (80 mR) and A IV (90 mR).


In Figure~\ref{compa}, we present the first comprehensive map of \ha\ emission in the region. The map is composed of over 1600 WHAM one-degree beams, processed similar to the WHAM Northern Sky Survey \citep{WHAMNSS}. The image is then constructed by integrating over velocities that sample Complex A ($-220$ to $-110$ \kms). First presented by \citet{DunHafMad05}, we have since doubled the total integration (to 120 s) for about half of the sightlines, mostly distributed along the Complex A chain. Due to considerable atmospheric line density at intensities $\sim 10$--100 mR, baseline fitting is the major source of uncertainty. Combining cleaned spectra obtained on multiple nights lessens the impact of any systematic contaminants. Nonetheless, the resulting map still shows very low contrast compared to the 21-cm map of the region, here from the Leiden-Dwingeloo survey (LDS, \citealp{HarBur97}). Most detected components have \ha\ intensities of 20--80 mR (EM $\sim$ 0.04--0.18 pc cm$^{-6}$, if $T_e = 8000$ K).


To study the relationship between the ionized and neutral components, we developed a simple, automated, Gaussian fitting algorithm to characterize positive detections in \ha\ and 21-cm spectra. Focusing on lines of sight that show a strong kinematic relationship between the two gas components, we identify directions in Figure~\ref{compa} (white crosses) where the fitted mean velocity of the \ha\ and 21-cm component differ by less than 10 \kms. Plots comparing the two emission spectra toward a representative sample of these directions are also shown. Exploring the data in the spectral domain adds considerable confidence that the WHAM data is tracing high-velocity gas, especially toward directions with measurable \hi\ column density. The correspondence between the neutral and ionized gas velocities seen in the selected spectra here continue in the larger statistical sample. The two phases are kinematically linked, as seen in other IVC and HVC studies (see \citealt{HafReyTuf01b} and \citealt{Haf05}). \ha\ intensity, on the other hand, does not correlate (or anti-correlate) with \hi\ column density. As with other studied complexes, the column density exhibits much larger variations over the extent of the HVC than the \ha\ intensity. Here, the neutral column spans about a factor of 10 to 20 while the emission measure varies by a factor of 3 to 5.

\section{Anti-Center Clouds}

\begin{figure}

\begin{center}
\includegraphics[width=9.5cm]{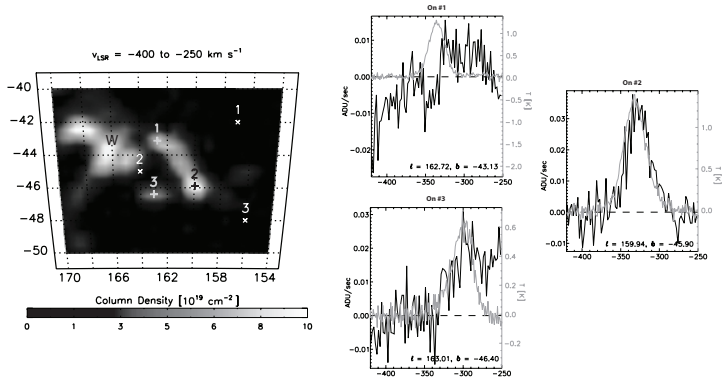}
\end{center}
\caption{
\label{accboth}
(\emph{left}) 21-cm emission from LDS in the direction of a few high-latitude Anti-Center HVCs. The location of the WHAM observations ($+$, ÒonÓ; $\times$, ÒoffÓ) are shown. The black ``W'' denotes the approximate location of the \ha\ detection (40 mR) reported by \citet{WeiVogWil01}. (\emph{right}) \ha\ spectra (black) with background subtracted towards the three ``on'' directions. The corresponding average 21-cm spectrum (gray) from LDS within the WHAM one-degree beam is also shown.
}

\end{figure}


\citet{PeePutMcK07} present 21-cm observations from Arecibo ($\sim 3\arcmin$) of a very high-velocity ($\vlsr \sim -330$ \kms) segment of the Anti-Center complex. Based on the morphology of the main feature and several red-shifted ($\Delta v \sim 30$ \kms), smaller clouds, they suggest that the gas is being dynamically shaped and stripped away by interaction with the halo. \citet{WeiVogWil01} report \ha\ emission of about 40 mR toward another cloud in the subcomplex (HVC 165-43-280; see Figure~\ref{accboth}). Based on this emission and a rough model of ionizing flux in the halo, they locate the HVC between about 10 and 20 kpc away. \citet{PeePutMcK07} find that their drag model predicts a consistent distance. 


Three ``on'' and three ``off'' directions (labeled in Figure~\ref{accboth}) were identified as the initial WHAM targets. Each location was observed for 120 s per exposure over two nights separated by a month, with ``on'' and ``off'' targets interleaved to aid atmospheric line removal. The total exposure time per direction is 1680 s. None of the ``off'' directions appear to contain Galactic emission, so a single, very high signal-to-noise ``off'' is created for each night and subtracted from each ``on'' direction. Results are shown in Figure~\ref{accboth} with the WHAM spectrum in black and the \hi\ spectrum from LDS shown in gray.


\ha\ emission is clearly detected in Directions 2 (60 mR) and 3 (35--90 mR) with velocity profiles similar to those of the 21-cm emission. Direction 1 shows no clear emission at the location of the \hi\ velocity. Although the baseline is far from flat, the range of intensities limits any possible emission to no more than one-third that of Direction 2 (20 mR). Direction 3 hints at red-shifted emission or a second component. This asymmetry is present in both nights when examined separately. Integrating over the whole velocity window gives a lower limit of 90 mR, while assuming a 30 \kms\ FWHM as measured in Direction 2 gives 35 mR for a component centered near $\vlsr = -290$ \kms. 





\end{document}